 \documentclass[twocolumn]{aastex631}

\shorttitle{Maximum ejecta masses in nova outbursts}
\shortauthors{I. Hachisu \& M. Kato}

\begin{document}

\title{Upper bound of ejecta mass in a nova outburst}


\author[0000-0002-0884-7404]{Izumi Hachisu}
\affil{Department of Earth Science and Astronomy,
College of Arts and Sciences, The University of Tokyo,
3-8-1 Komaba, Meguro-ku, Tokyo 153-8902, Japan}
\email{izumi.hachisu@outlook.jp}

\author[0000-0002-8522-8033]{Mariko Kato}
\affil{Department of Astronomy, Keio University,
Hiyoshi, Kouhoku-ku, Yokohama 223-8521, Japan}




.


\begin{abstract}
We present the maximum ejecta mass $(M_{\rm ej})_{\rm max}$
and the maximum ratio of ejecta mass and accreted mass 
$(M_{\rm ej}/M_{\rm acc})_{\rm max}$ of a nova for various
white dwarf (WD) masses ($M_{\rm WD}=0.6$ - 1.38 $M_\sun$) and mass accretion
rates ($\dot{M}_{\rm acc}=1\times 10^{-11}$ - $3\times 10^{-7} ~M_\sun$ 
yr$^{-1}$) based on the energy balance with nuclear burning.  These maximum
values serve as an upper bound of mass ejection for individual novae.  
Recently, B. E. Schaefer concluded that the WD masses in the recurrent novae
U Sco and T CrB decreased at nova explosions, because the ejected mass is
much larger than the accreted mass, i.e., $M_{\rm ej}/M_{\rm acc}= 26$ and
$540$, respectively.  These values are derived from the orbital period
change at the nova explosions.  Recurrent novae have been considered to be
a progenitor system of Type Ia supernovae (SNe Ia) because their WD masses
are now close to, and will possibly grow up to, 1.38 $M_\sun$ at which WDs
explode as SNe Ia. From the different view point of energy generation at the
thermonuclear runaway, we have obtained the much smaller value of the maximum
ratio of $M_{\rm ej}/M_{\rm acc}\lesssim 2.6$ for a $1.37 ~M_\sun$ WD.  This
conclusion simply means that the nuclear (hydrogen) burning cannot release
energy enough to expel such a large ejecta mass as B. E. Schaefer's claims.
We also conclude that $(M_{\rm ej}/M_{\rm acc})_{\rm max}$ hardly increases
even if we include the effect of frictional mass ejection process
in the common envelope phase of a nova.
\end{abstract}





\section{Introduction}
\label{introduction}

It has long been suggested that recurrent novae are a progenitor system 
of Type Ia supernovae (SNe Ia) \citep[e.g., ][]{hkn96, hknu99,
hkn99, hkn10, hac01kb, han04, kat12review, li97, mao14}.

Very recently, \citet{schaefer25usco} and \citet{schaefer25tcrb,
schaefer26tpyx} obtained the ratios of the ejecta mass ($M_{\rm ej}$) 
and accreted mass ($M_{\rm acc}$) in the recurrent novae U Sco, T CrB,
and T Pyx to be $M_{\rm ej}/M_{\rm acc}=26$, $540$, and
$\gg 11.3$, respectively.  These very large ratios of 
$M_{\rm ej}/M_{\rm acc}$ are derived from the combination of
the orbital period change and orbital angular momentum loss by the ejecta.
If that is the case, the white dwarf (WD) loses its mass much more
than the accreted mass.  As a result, the WD mass is decreasing
and cannot grow to $1.38 ~M_\sun$ at which a carbon-oxygen (CO) WD
explodes as an SN Ia.

In the present paper, we examine this problem from a different view point
of required energy to eject such a large mass in a nova outburst.
Section \ref{maximum_mass_ejecta} describes our estimate of maximum ejecta
mass in a nova outburst based on the energy balance with hydrogen burning.
Then, we compare our results with B. E. Schaefer's values.
We also examine a possible effect of frictional mass ejection process
in the common envelope phase of a nova in Section 
\ref{frictional_mass_ejection}.  
Discussion and conclusions follow in Sections 
\ref{sec_discussion} and \ref{sec_conclusion}, respectively.


\begin{figure*}
\gridline{\fig{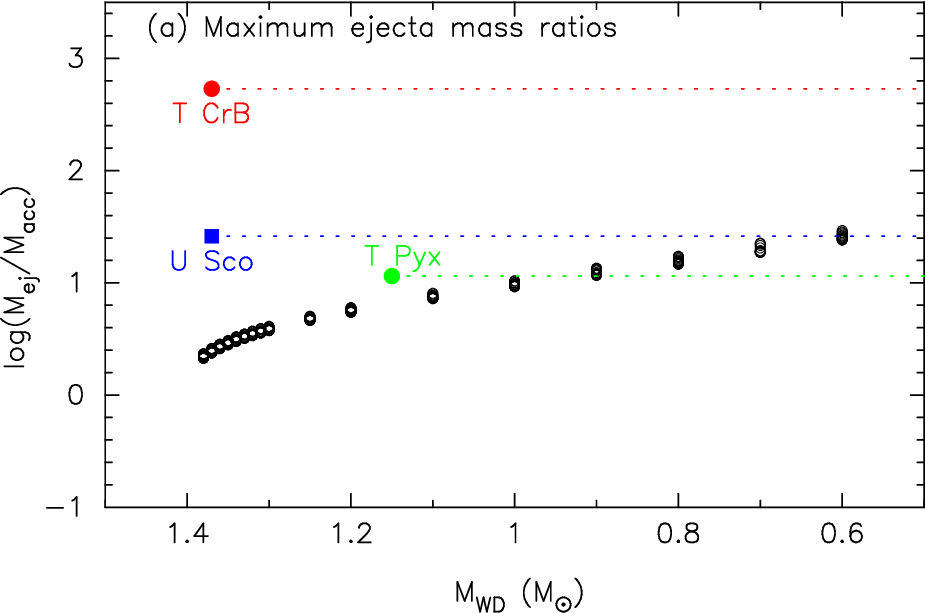}{0.5\textwidth}{}
          }
\gridline{
          \fig{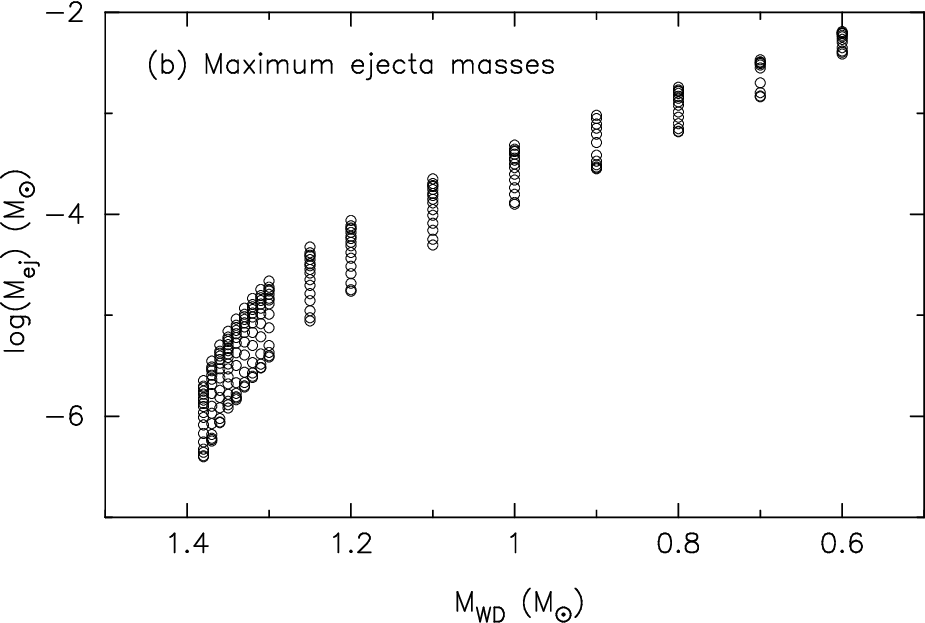}{0.5\textwidth}{}
          }
\caption{
(a) The ratio of the maximum ejecta mass and accreted mass $(M_{\rm ej}/
M_{\rm acc})_{\rm max}$ is plotted for various WD masses $M_{\rm WD}$ and
mass accretion rates $\dot{M}_{\rm acc}$ on to the WD by open black circles.
These maximum values serve as an upper bound for individual novae.
These data are tabulated in Table \ref{table_max_ejecta_masses} 
of Appendix \ref{table_maximum_ejecta}.  We add the positions of three
recurrent novae U Sco (filled blue square; $M_{\rm ej}/M_{\rm acc}=26$),
T CrB (filled red circle; $M_{\rm ej}/M_{\rm acc}=540$),
and T Pyx (filled green circle; $M_{\rm ej}/M_{\rm acc}\gg 11.3$),
taken from \citet{schaefer25usco}, \citet{schaefer25tcrb}, 
and \citet{schaefer26tpyx}, respectively.
(b) The maximum ejecta mass $(M_{\rm ej})_{\rm max}$ is plotted for various
WD masses and mass accretion rates by open black circles.  The data are
also tabulated in Table \ref{table_max_ejecta_masses} of Appendix
\ref{table_maximum_ejecta}.
\label{max_ratio_ejecta}}
\end{figure*}


\begin{figure*}
\gridline{\fig{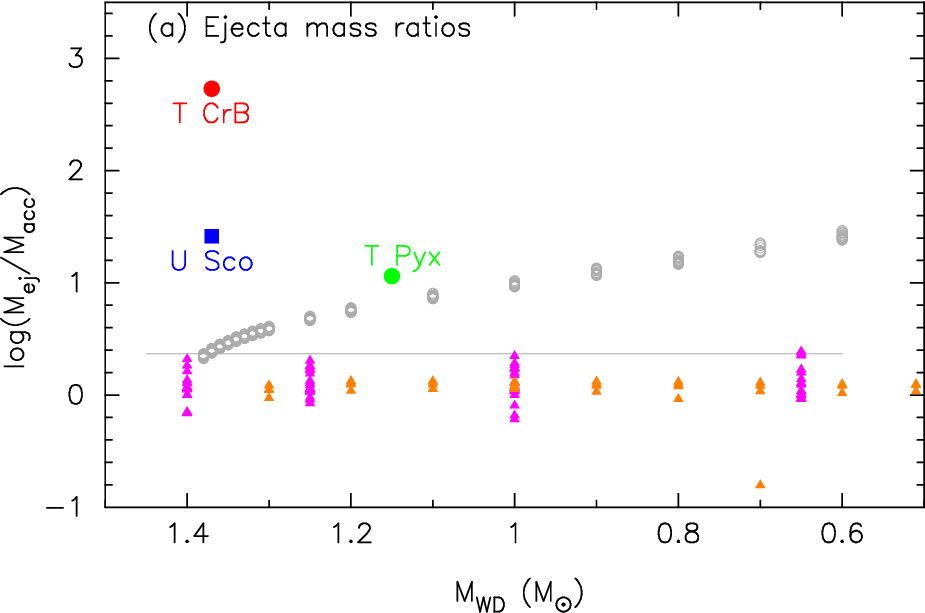}{0.5\textwidth}{}
          }
\gridline{
          \fig{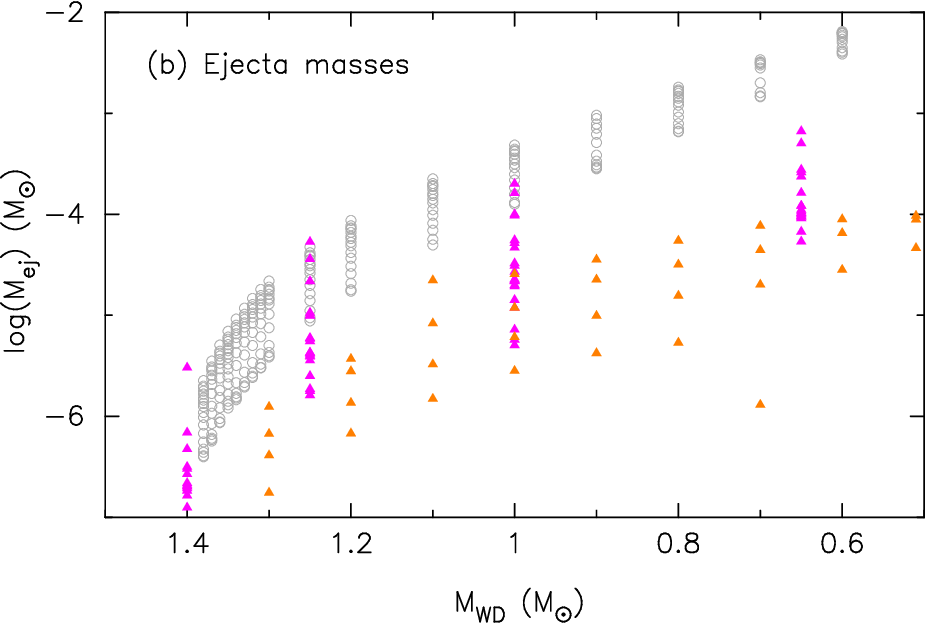}{0.5\textwidth}{}
          }
\caption{
(a) Same as those in Figure \ref{max_ratio_ejecta}(a), but we add the 
individual $M_{\rm ej}/M_{\rm acc}$ results of \citet{yar05}
by filled magenta triangles and of \citet{chen19} by filled orange triangles.
Note that these are the direct results of nova outburst calculations. 
Our maximum value serves as an upper bound for these individual calculations.
The horizontal gray line indicates the value of $(M_{\rm ej}/
M_{\rm acc})_{\rm max}= 2.33$ for the case of $M_{\rm WD}=1.38 ~M_\sun$
and $\dot{M}_{\rm acc}= 2.7\times 10^{-7} ~M_\sun$ yr$^{-1}$.
(b) Same as those in Figure \ref{max_ratio_ejecta}(b),
but we add the direct nova outburst calculation results
of \citet{yar05} by filled magenta triangles and
of \citet{chen19} by filled orange triangles. 
\label{max_ejecta_masses_yaron}}
\end{figure*}

\section{Estimate of maximum ejecta masses from shell flash calculations}
\label{maximum_mass_ejecta}

A nova is a thermonuclear runaway event on a mass-accreting WD.  When the
mass of the hydrogen-rich envelope reaches a critical value ($M_{\rm ig}$),
unstable hydrogen burning ignites to trigger a thermonuclear runaway
\citep[e.g.,][]{spa78, sio79, nar80, ibe82, pri95k}.  When the photosphere
of the hydrogen-rich envelope expands to reach $\sim 0.1 ~R_\sun$, 
optically thick winds emerge from the photosphere
\citep[e.g., ][]{kat94h, kat22sha, kat25hs}.
Thus, the hydrogen-rich envelope loses its mass by winds. 

\subsection{Ignition mass}
\label{ignition_mass}
When the mass of a hydrogen-rich envelope 
on the accreting WD increases and reaches
a critical value of $M_{\rm ig}$, unstable hydrogen burning ignites 
at the bottom of the envelope.  We define the mass of the envelope 
at ignition as the ignition mass $M_{\rm ig}$.  The ignition mass
depends on the WD mass $M_{\rm WD}$ and mass accretion rate
$\dot{M}_{\rm acc}$ on to the WD.  
\citet{kat14shn} and \citet{hac20skhs} have already obtained the ignition
masses based on a Henyey type evolution code, but presented them nowhere.
We here tabulate the ignition masses for various WD
masses and mass accretion rates 
in Table \ref{table_max_ejecta_masses} 
of Appendix \ref{table_maximum_ejecta}.

\subsection{Ejecta mass}
\label{ejecta_mass}

We estimate the maximum (possible) ejecta mass $(M_{\rm ej})_{\rm max}$
by a simple energy budget:
The energy to eject $M_{\rm ej}$ into interstellar space is given by
\begin{equation} 
E_{\rm eject} = 2\phi_{\rm G} M_{\rm ej},
\label{ejection_energy}
\end{equation} 
where $E_{\rm eject}$ is the energy to eject the envelope mass of
$M_{\rm ej}$ and 
\begin{equation} 
\phi_{\rm G}={{GM_{\rm WD}}\over {R_{\rm b}}}
\end{equation} 
is the gravitational potential at the bottom radius $R_{\rm b}$
of the hydrogen-rich envelope.
A factor of 2 is required for the envelope thermal energy to heat and expand
the envelope, which is the same amount of, or slightly larger than, 
the gravitational energy. 
See Figure 2 in \cite{kat83}, and Table 3 and Figure 13 in \citet{kat22sha}
for more details. 

This amount of energy is supplied by nuclear burning.  Here, we assume that
all the hydrogen in the envelope is burned into helium, and obtain the
maximum ejecta mass to be
\begin{equation} 
(M_{\rm ej})_{\rm max}= {{E_{\rm eject}} \over {2\phi_{\rm G}}}
= {{ X \varepsilon_{\rm n} M_{\rm ig}} \over {2\phi_{\rm G}}},
\label{eq_maximum_ejecta_mass}
\end{equation} 
where $X=0.7$ is the hydrogen content of the accreted matter by mass
and $\varepsilon_{\rm n}= \Delta m c^2 = 0.00715 c^2= 6.426\times 10^{18}$
erg g$^{-1}$ is the hydrogen burning energy into helium.
Thus, we obtain the maximum ejecta mass ratio of
\begin{equation} 
\left({{M_{\rm ej}} \over {M_{\rm ig}}}\right)_{\rm max} 
= {{X \varepsilon_{\rm n}  R_{\rm b}} \over {2 G M_{\rm WD} }},
\label{max_ejecta_mass_ratio}
\end{equation} 
where we substitute $\phi_{\rm G}= G M_{\rm WD} / R_{\rm b}$ 
into Equation (\ref{eq_maximum_ejecta_mass}).

It should be noted that the ignition mass $M_{\rm ig}$ is slightly
larger than the accreted mass $M_{\rm acc}$ because there is a small
leftover hydrogen-rich envelope mass in the previous outburst.
However, if all the hydrogen is burned into helium (our assumption),
there is no leftover of hydrogen-rich envelope.  
Therefore, in what follows, we regard that
\begin{equation} 
M_{\rm ig}=M_{\rm acc},
\label{mig_equal_macc}
\end{equation} 
and
\begin{equation} 
\left({{M_{\rm ej}} \over {M_{\rm acc}}}\right)_{\rm max} 
= {{X \varepsilon_{\rm n}  R_{\rm b}} \over {2 G M_{\rm WD} }}.
\label{max_ratio_ej_acc}
\end{equation}

\subsection{Ratio of ejecta mass and accreted mass}
\label{ratio_ejecta_accreted_mass}

Equation (\ref{max_ratio_ej_acc}) simply shows  
that the ratio of the maximum ejecta mass and accreted mass,
$(M_{\rm ej}/ M_{\rm acc})_{\rm max}$, is determined by 
the WD mass $M_{\rm WD}$ and radius $R_{\rm b}$. 
We tabulate the  radius $R_{\rm b}$ in Table \ref{table_max_ejecta_masses}  
of Appendix \ref{table_maximum_ejecta} taken from our calculation, 
and obtained $(M_{\rm ej}/ M_{\rm acc})_{\rm max}=(M_{\rm ej}/
M_{\rm ig})_{\rm max}$ as well as $(M_{\rm ej})_{\rm max}$. 
Figure \ref{max_ratio_ejecta} shows (a) the maximum ratio $(M_{\rm ej}/
M_{\rm acc} )_{\rm max}$ and (b) the maximum ejecta mass 
$(M_{\rm ej})_{\rm max}$ against the WD mass $M_{\rm WD}$
for various mass accretion rates. 

The radius $R_{\rm b}$ is almost determined by the WD mass $M_{\rm WD}$,
although it slightly depends on the mass accretion rate $\dot{M}_{\rm acc}$.
Therefore, the ratio $(M_{\rm ej}/ M_{\rm acc})_{\rm max}$
is close to a unique function of the WD mass $M_{\rm WD}$ as shown
in Figure \ref{max_ratio_ejecta}(a).    
The ratio $\log(M_{\rm ej}/ M_{\rm acc})_{\rm max}$ decreases linearly
with the WD mass increasing but rapidly drops toward 
$M_{\rm WD}= 1.38 ~M_\sun$, where $(M_{\rm ej}/ M_{\rm acc})_{\rm max}
=(M_{\rm ej}/ M_{\rm ig})_{\rm max} \approx 2.1$ - 2.3, 
because the WD radius becomes smaller toward the Chandrasekhar mass limit.

\citet{schaefer25usco} and \citet{schaefer25tcrb} obtained the ratios of 
 $M_{\rm ej}/ M_{\rm acc} = 26$ for U Sco (filled blue square) and
$M_{\rm ej}/ M_{\rm acc} = 540$ for T CrB (filled red circle),
respectively, as shown in Figure \ref{max_ratio_ejecta}(a). 

\citet{hac00kkm} and \citet{hac99ktcrb, hac01kb} estimated 
the WD masses of U Sco and T CrB to be $\sim 1.37 ~M_\sun$ 
based on their model light curve fittings.  
See also $M_{\rm WD}=1.55\pm 0.24 ~M_\sun$ and $M_2=0.88\pm 0.17 ~M_\sun$
for U Sco observationally estimated by \citet{thorou01},
and $M_{\rm WD}=1.37\pm 0.01 ~M_\sun$ and $M_2=0.69^{+0.02}_{-0.01} ~M_\sun$
for T CrB by \citet{hink25}.  Here, $M_2$ is the mass of a companion star
to the WD star.

As in Table \ref{table_max_ejecta_masses} 
of Appendix \ref{table_maximum_ejecta}, the WD radius is $\log R_{\rm b}/
R_\sun = -2.563$ for the $1.37 ~M_\sun$ WD with $\dot{M}_{\rm acc}= 
1\times 10^{-11} ~M_\sun$ yr$^{-1}$.  Then, we have $(M_{\rm ej}/ 
M_{\rm acc})_{\rm max}= 2.34$ from Equation (\ref{max_ratio_ej_acc}).
If we adopt $\dot{M}_{\rm acc}= 1\times 10^{-7} ~M_\sun$ yr$^{-1}$,
we have $(M_{\rm ej}/M_{\rm acc})_{\rm max} = 2.54$. 
Our results do not support B. E. Schaefer's large values. 
This conclusion simply says that, even if all the hydrogen burns
into helium, the nuclear burning energy is not enough to expel
such a large ejecta mass of $M_{\rm ej}/ M_{\rm acc} \gtrsim 2.6$
in our $1.37 ~M_\sun$ WD model.

\subsection{T Pyx}
\label{discussion_t_pyx}

Very recently, \citet{schaefer26tpyx} obtained the ejecta mass ratio
of $M_{\rm ej}/M_{\rm acc}\gg 11.3$ for the recurrent nova T Pyx.
We add this case to Figure \ref{max_ratio_ejecta}(a) by the filled green
circle, adopting $M_{\rm WD}= 1.15 ~M_\sun$ for T Pyx after \citet{hac21k}
who obtained the WD mass by a light curve fitting with the decay phase
in the 2011 outburst.  Our upper bound is $M_{\rm ej}/M_{\rm acc} \lesssim
(M_{\rm ej}/M_{\rm acc})_{\rm max}=$6.25 - 6.79.

However, if the WD mass of T Pyx is less massive than $0.9 ~M_\sun$,
B. E. Schaefer's lowest limit $M_{\rm ej}/M_{\rm acc}=11.3$ 
is acceptable with our upper bound as suggested by the horizontal green
dotted line.  \citet{uthas10ks} obtained the WD mass of $M_{\rm WD}= 0.7\pm
0.2 ~M_\sun$ and the companion mass of $M_2=0.14 \pm 0.03 ~M_\sun$ for the
T Pyx system.  If it is the case, B. E. Schaefer's ejecta mass ratio of
$M_{\rm ej}/M_{\rm acc}=11.3$ are not rejected at least with our upper bound
from the energetics point of view.

On the other hand, the shortest recurrence period of T Pyx,
$t_{\rm rec}= 7.49$ yr \citep[see, e.g., Table 2 of ][]{schaefer26tpyx},
can be realized only for more massive WDs than $M_{\rm WD}\gtrsim 1.15
~M_\sun$ \citep[see, e.g., Figure 6 of ][]{kat14shn}.  
We may conclude that B. E. Schaefer's result of $M_{\rm ej}/M_{\rm acc}
\gg 11.3$ is not supported.

\subsection{Comparison with other nova calculations}
\label{comparison_nova_calculations}

In this subsection, we compare our results with other nova calculations.
\citet{yar05} calculated nova outburst evolutions for various WD masses
and mass accretion rates and tabulated their results on the $M_{\rm ej}$
and $M_{\rm acc}$.  Figure \ref{max_ejecta_masses_yaron}(a) compares our
maximum ratios with \citet{yar05}'s ratios (filled magenta triangles).
Figure \ref{max_ejecta_masses_yaron}(b) also compares our maximum ejecta
masses with their ejecta masses (filled magenta triangles).  Here, we take 
their ejecta masses $M_{\rm ej}$ and accreted masses $M_{\rm acc}$
from their Table 2 and then obtain their ratios $M_{\rm ej}/M_{\rm acc}$.
Note that the ejecta mass possibly exceeds the accreted mass
because they includes hydrogen diffusion into the WD core in the quiescent 
phase and thermonuclear runaway occurs somewhat below the accreted layer. 
Our maximum ejecta mass and mass ratios (gray symbols in Figure
\ref{max_ejecta_masses_yaron}) serve as an upper bound for 
\citet{yar05}'s calculations.  

In Figure \ref{max_ejecta_masses_yaron}(a) and (b),
we also add the results for the case of $Z=0.02$
of \citet{chen19} who calculated nova outbursts for various WD masses and
mass accretion rates including some mixing process in the nova envelope
(their Table 5).

Figure \ref{max_ejecta_masses_yaron}(a) shows that all of the ratios from
\citet{yar05} (magenta triangles) and \citet{chen19} (orange triangles)
are below our ratios of $(M_{\rm ej}/ M_{\rm acc})_{\rm max}$
(open gray circles).  This means that our simple energetics are
consistent with their numerical calculations.

\section{Frictional mass ejection}
\label{frictional_mass_ejection}

Some recent nova explosion calculations included frictional angular
momentum loss (FAML) as a main orbital angular momentum loss of the binaries
\citep[e.g.,][]{cho21ms, spa21s, shen22q}.  This suggests that some part
of the ejection energy is supplied by the frictional mass ejection process.  

If the frictional mass ejection effectively works, the envelope matter 
expands to over the binary orbit (not infinite) 
and then will be accelerated by frictional
energy input toward infinite.  This certainly requires 
smaller energy than $2 G M_{\rm WD}/R_b$ in Equation
(\ref{ejection_energy}).

We need the energy of
\begin{equation} 
E_{\rm expansion} = 2\left( {{G M_{\rm WD}}\over{R_{\rm b}}}
- {{G M_{\rm WD}}\over{a_{\rm orb}}} \right) M_{\rm ej},
\label{energy_expand_orbit}
\end{equation} 
when the envelope mass (ejecta mass $M_{\rm ej}$) expands to the binary
orbit.  Here, $a_{\rm orb}$ is the orbital radius.   
If we define 
\begin{equation} 
\beta \equiv 1- {{R_{\rm b}} \over {a_{\rm orb}}},
\label{effect_expand_orbit}
\end{equation} 
we obtain
\begin{equation} 
E_{\rm expansion} = \beta E_{\rm eject}.
\label{energy_ratio_orbit}
\end{equation} 
Therefore, the upper bound of ejecta mass could increase by a factor
of $1/\beta$, that is,
$(M_{\rm ej}/M_{\rm acc})_{\rm max} \lesssim 2.6/\beta$
for a $1.37 ~M_\sun$ WD.  

For U Sco, we have $\beta = 1 - 0.002735 R_\sun / 6.5 R_\sun = 0.99958$,
where the WD radius $\log R_{\rm b}/ R_\sun = -2.563$ 
for the $1.37 ~M_\sun$ WD is taken from Table \ref{table_max_ejecta_masses} 
of Appendix \ref{table_maximum_ejecta}, and $a_{\rm orb}=6.5 ~R_\sun$ is
taken from \citet{thorou01}.
Similarly, we have $\beta = 1 - 0.002735 R_\sun / 199.5 R_\sun = 0.999986$
for T CrB, where $a_{\rm orb}=199.5 ~R_\sun$ is calculated from the
results of \citet{hink25}.  We also obtain
$\beta = 1 - 0.006081 R_\sun / 0.8254 R_\sun = 0.99263$
for T Pyx, where the WD radius $\log R_{\rm b}/ R_\sun = -2.216$ 
for the $1.15 ~M_\sun$ WD with $\dot{M}_{\rm acc}=1\times 10^{-11} ~M_\sun$
yr$^{-1}$ is calculated from an interpolation between 1.1 and 1.2 $M_\sun$
WDs (Table \ref{table_max_ejecta_masses} 
of Appendix \ref{table_maximum_ejecta}), and 
$a_{\rm orb}=0.8254 ~R_\sun$ is calculated from $M_{\rm WD}=1.15 ~M_\sun$
\citep{hac21k}, $M_2=0.15 ~M_\sun$, and $P_{\rm orb}=0.076229$ day
\citep{schaefer26tpyx}.  Thus, the upper bound of ejecta mass is not
essentially affected by the effect of frictional mass ejection because
$\beta\approx 1$ for U Sco, T CrB, and T Pyx. 


\section{Discussion}
\label{sec_discussion}

We have shown that our analysis on the upper bound of ejecta mass in a nova
outburst does clearly not support B. E. Schaefer's estimates of
$M_{\rm ej}/M_{\rm acc}=26$, 540, and $\gg 11.3$, in U Sco, T CrB, and T Pyx,
respectively, because $(M_{\rm ej}/M_{\rm acc})_{\rm max}\lesssim 2.6$
for $M_{\rm WD}=1.37 ~M_\sun$ (U Sco and T CrB) and 
$(M_{\rm ej}/M_{\rm acc})_{\rm max}\lesssim 6.8$ for
$M_{\rm WD}=1.15 ~M_\sun$ (T Pyx).  

We are not confident of the reason why B. E. Schaefer derived such
large values of the ejecta masses, but suggest possible reasons.

\noindent
(1) Other angular-momentum-loss mechanisms:\\
B. E. Schaefer converts orbital
period change to $M_{\rm ej}/M_{\rm acc}$ using a specific assumption;
all the angular momentum loss is owing to the nova ejecta.  
When the ejecta collides with the companion star, a part of the envelope
mass of the companion could be stripped off \citep[see, e.g.,][]{hkn99}.
If the mass of stripped matter is comparable with the ejecta mass,
its specific orbital angular momentum could be much larger than that of
the nova ejecta themselves.  This effect causes the orbital period change
independently of the nova ejecta mass.
The trend of increasing $M_{\rm ej}/M_{\rm acc}=11.3$, 26, and 540 
corresponds to the increasing orbital radius of 0.83,
6.5, 199.5 $R_\sun$, suggesting looser coupling with the gravity of the
companion envelope, in the order of T Pyx, U Sco, and T CrB.
However, B. E. Schaefer apparently attributes all of the period change
$\Delta P/P$ to the nova ejecta. 

\noindent
(2) Angular-momentum-loss geometry: \\
 As explained in above (1), ejection of the stripped companion
matter could be non-spherical and slower than that of the nova ejecta.
Its angular momentum coupling with orbital
motion is different from B. E. Schaefer's conversion
\citep[see, e.g.,][]{hkn99}. 

\noindent
(3) Period measurement precision:\\
In particular, $M_{\rm ej}/M_{\rm acc}=540$ of T CrB suggests
that the stripped mass could be much larger than the nova ejecta mass
as explained in above (1).
Such a large value comes from the period change $\Delta P/P$ of the
binary, and this estimate depends on the $O-C$ values before the 1946
eruption, especially several data before 1910 \citep[particular two data
in 1867-1871 and 1874-1880, ][]{schaefer25tcrb}.
However, they are rather scattered than the $O-C$ data after 1946.
We suppose that even tiny systematic errors in eclipse timing
before 1910 could cause a large error of $\Delta P/P$.

\section{Conclusions}
\label{sec_conclusion}

We have estimated the maximum ejecta mass ratios of nova outbursts, 
$(M_{\rm ej}/M_{\rm acc})_{\rm max}$, for various WD masses and mass
accretion rates based on the energetics of hydrogen burning on the WD.
These maximum values serve as an upper bound for individual novae.
Here, $M_{\rm ej}$ is the ejecta mass and $M_{\rm acc}$ is the accreted
hydrogen-rich envelope mass.
Our results are summarized as follows: \\
\begin{enumerate}
\item Our $1.37 ~M_\sun$ WD model calculations with the mass-accretion rates
between $1\times 10^{-11}$ and $3\times 10^{-7} ~M_\sun$ yr$^{-1}$ show
that the upper bounds for the ejecta mass ratios $(M_{\rm ej}/
M_{\rm acc})_{\rm max}$ are between 2.37 and 2.6.  These results are much
smaller than, and therefore do not support, \citet{schaefer25usco}'s 
$M_{\rm ej}/M_{\rm acc}=26$ for U Sco and \citet{schaefer25tcrb}'s 
$M_{\rm ej}/M_{\rm acc}=540$ for T CrB.
\item Our $1.15 ~M_\sun$ WD model calculations 
with the mass-accretion rates between
$1\times 10^{-11}$ and $1.6\times 10^{-7} ~M_\sun$ yr$^{-1}$, which are
interpolated from our $1.1$ and $1.2 ~M_\sun$ WDs, show that the upper 
bounds for the maximum ejecta mass ratios $(M_{\rm ej}/M_{\rm acc})_{\rm 
max}$ are between 6.25 and 6.79.  These results also do not support 
\citet{schaefer26tpyx}'s $M_{\rm ej}/M_{\rm acc}\gg 11.3$ for T Pyx.
\item We examine a possibility that frictional mass ejection in the common
envelope phase plays a role.
Even if we include such an effect, energy required by expansion of the
envelope to the binary orbit is nearly equal to that without frictional
mass ejection.  Therefore, our requirements for upper bound is not
essentially changed.
\item 
We do not agree with B. E. Schaefer's conclusion that recurrent novae
do not evolve to Type Ia supernova progenitors, because it is resulted
from the $M_{\rm ej}/M_{\rm acc}$ of the above three recurrent novae. 
\end{enumerate}

\begin{acknowledgments}
We are grateful to the anonymous referee for valuable comments. 
\end{acknowledgments}



\appendix
\section{Maximum ejecta masses in nova outbursts}
\label{table_maximum_ejecta}

We tabulate our numerical results on the maximum ejecta mass ratios 
of $(M_{\rm ej}/M_{\rm ig})_{\rm max}$ and maximum ejecta masses of 
$(M_{\rm ej})_{\rm max}$ for various white dwarf (WD) masses $M_{\rm WD}$
and mass accretion rates $\dot{M}_{\rm acc}$
in Table \ref{table_max_ejecta_masses}.
In our calculation of the maximum ejecta mass, we assume that
$M_{\rm ig}=M_{\rm acc}$ because there is no leftover hydrogen-rich
matter in the previous outburst.

\startlongtable
\begin{deluxetable}{lllrcl}
\tabletypesize{\scriptsize}
\tablecaption{Theoretical maxima of ejecta masses
\label{table_max_ejecta_masses}}
\tablehead{\colhead{$M_{\rm WD}$} &
\colhead{$\dot{M}_{\rm acc}$} &
\colhead{$M_{\rm ig}$} &
\colhead{$\log(R_b)$} &
\colhead{$(M_{\rm ej}/M_{\rm ig})_{\rm max}$} &
\colhead{$(M_{\rm ej})_{\rm max}$} \\
\colhead{($M_\sun$)} &
\colhead{($M_\sun$ yr$^{-1}$)} &
\colhead{($M_\sun$)} &
\colhead{($R_\sun$)} &
\colhead{}&
\colhead{($M_\sun$)} 
}
\startdata
0.60 &  1.0E-11  &    0.000269  &  -1.915 &          24 &    0.00645  \\ 
0.60 &  3.0E-11  &    0.000258  &  -1.911 &        24.2 &    0.00627  \\ 
0.60 &  5.0E-11  &    0.000255  &  -1.909 &        24.4 &    0.00621  \\ 
0.60 &  1.0E-10  &    0.000254  &  -1.906 &        24.5 &    0.00622  \\ 
0.60 &  3.0E-10  &    0.000248  &  -1.900 &        24.9 &    0.00616  \\ 
0.60 &  1.0E-9  &    0.000233  &  -1.891 &        25.4 &    0.00593  \\ 
0.60 &  1.6E-9  &    0.000223  &  -1.887 &        25.6 &    0.00572  \\ 
0.60 &  3.0E-9  &    0.000207  &  -1.879 &        26.1 &     0.0054  \\ 
0.60 &  5.0E-9  &    0.000191  &  -1.872 &        26.5 &    0.00506  \\ 
0.60 &  1.0E-8  &    0.000164  &  -1.860 &        27.3 &    0.00446  \\ 
0.60 &  1.6E-8  &    0.000148  &  -1.849 &          28 &    0.00414  \\ 
0.60 &  2.0E-8  &    0.000142  &  -1.843 &        28.4 &    0.00402  \\ 
0.60 &  3.0E-8  &    0.000132  &  -1.829 &        29.3 &    0.00385  \\ 
0.70 &  1.0E-11  &    0.000182  &  -1.959 &        18.6 &    0.00339  \\ 
0.70 &  3.0E-11  &    0.000173  &  -1.957 &        18.7 &    0.00323  \\ 
0.70 &  5.0E-11  &    0.000169  &  -1.955 &        18.8 &    0.00318  \\ 
0.70 &  1.0E-10  &    0.000165  &  -1.953 &        18.9 &    0.00311  \\ 
0.70 &  3.0E-10  &    0.000157  &  -1.948 &        19.1 &      0.003  \\ 
0.70 &  1.0E-09  &    0.000145  &  -1.941 &        19.4 &    0.00281  \\ 
0.70 &  1.0E-08  &    9.78E-05  &  -1.918 &        20.5 &      0.002  \\ 
0.70 &  3.0E-08  &    7.44E-05  &  -1.896 &        21.5 &     0.0016  \\ 
0.70 &  5.0E-08  &    6.66E-05  &  -1.881 &        22.3 &    0.00148  \\ 
0.70 &  6.0E-08  &    6.44E-05  &  -1.875 &        22.6 &    0.00145  \\ 
0.80 &  1.0E-11  &    0.000124  &  -2.005 &        14.6 &    0.00181  \\ 
0.80 &  3.0E-11  &    0.000116  &  -2.003 &        14.7 &     0.0017  \\ 
0.80 &  5.0E-11  &    0.000113  &  -2.001 &        14.8 &    0.00166  \\ 
0.80 &  1.0E-10  &    0.000111  &  -1.999 &        14.9 &    0.00165  \\ 
0.80 &  3.0E-10  &    0.000105  &  -1.996 &          15 &    0.00158  \\ 
0.80 &  1.0E-9  &    9.71E-05  &  -1.990 &        15.2 &    0.00147  \\ 
0.80 &  1.6E-9  &    9.22E-05  &  -1.988 &        15.2 &     0.0014  \\ 
0.80 &  3.0E-9  &    8.42E-05  &  -1.983 &        15.4 &     0.0013  \\ 
0.80 &  5.0E-9  &    7.74E-05  &  -1.979 &        15.6 &     0.0012  \\ 
0.80 &  1.0E-8  &    6.53E-05  &  -1.973 &        15.8 &    0.00103  \\ 
0.80 &  1.6E-8  &    5.71E-05  &  -1.967 &          16 &   0.000914  \\ 
0.80 &  3.0E-8  &    4.82E-05  &  -1.956 &        16.4 &    0.00079  \\ 
0.80 &  5.0E-8  &    4.19E-05  &  -1.945 &        16.8 &   0.000705  \\ 
0.80 &  7.0E-8  &    3.86E-05  &  -1.936 &        17.2 &   0.000663  \\ 
0.80 &  7.5E-8  &     3.8E-05  &  -1.934 &        17.2 &   0.000656  \\ 
0.90 &  1.0E-11  &    8.22E-05  &  -2.054 &        11.6 &   0.000956  \\ 
0.90 &  3.0E-11  &    7.58E-05  &  -2.052 &        11.7 &   0.000886  \\ 
0.90 &  3.0E-10  &     6.6E-05  &  -2.047 &        11.8 &    0.00078  \\ 
0.90 &  1.0E-9  &    5.95E-05  &  -2.042 &          12 &   0.000712  \\ 
0.90 &  3.0E-9  &    5.14E-05  &  -2.037 &        12.1 &   0.000621  \\ 
0.90 &  1.0E-8  &    4.16E-05  &  -2.028 &        12.3 &   0.000514  \\ 
0.90 &  3.0E-8  &    3.02E-05  &  -2.014 &        12.8 &   0.000385  \\ 
0.90 &  5.0E-8  &    2.59E-05  &  -2.006 &          13 &   0.000336  \\ 
0.90 &  7.0E-8  &    2.35E-05  &  -2.000 &        13.2 &    0.00031  \\ 
0.90 &  9.0E-8  &     2.2E-05  &  -1.994 &        13.4 &   0.000293  \\ 
0.90 &  1.0E-7  &    2.14E-05  &  -1.991 &        13.4 &   0.000287  \\ 
0.90 &  1.1E-7  &    2.09E-05  &  -1.989 &        13.5 &   0.000282  \\ 
1.00 &  1.0E-11  &    5.25E-05  &  -2.109 &        9.22 &   0.000485  \\ 
1.00 &  3.0E-11  &    4.77E-05  &  -2.107 &        9.26 &   0.000442  \\ 
1.00 &  5.0E-11  &    4.59E-05  &  -2.106 &        9.29 &   0.000426  \\ 
1.00 &  1.0E-10  &    4.48E-05  &  -2.105 &        9.31 &   0.000417  \\ 
1.00 &  3.0E-10  &    4.19E-05  &  -2.102 &        9.37 &   0.000392  \\ 
1.00 &  1.0E-9  &    3.81E-05  &  -2.099 &        9.44 &    0.00036  \\ 
1.00 &  1.6E-9  &    3.63E-05  &  -2.097 &        9.48 &   0.000344  \\ 
1.00 &  3.0E-9  &     3.3E-05  &  -2.094 &        9.55 &   0.000315  \\ 
1.00 &  5.0E-9  &    3.03E-05  &  -2.092 &        9.59 &    0.00029  \\ 
1.00 &  1.0E-8  &    2.57E-05  &  -2.087 &         9.7 &   0.000249  \\ 
1.00 &  1.6E-8  &    2.23E-05  &  -2.084 &        9.77 &   0.000218  \\ 
1.00 &  3.0E-8  &    1.85E-05  &  -2.077 &        9.93 &   0.000184  \\ 
1.00 &  5.0E-8  &    1.57E-05  &  -2.071 &        10.1 &   0.000158  \\ 
1.00 &  1.0E-7  &    1.27E-05  &  -2.059 &        10.3 &   0.000131  \\ 
1.00 &  1.2E-7  &     1.2E-05  &  -2.055 &        10.4 &   0.000126  \\ 
1.10 &  1.0E-11  &    3.11E-05  &  -2.174 &        7.22 &   0.000225  \\ 
1.10 &  3.0E-11  &    2.78E-05  &  -2.172 &        7.25 &   0.000202  \\ 
1.10 &  5.0E-11  &    2.66E-05  &  -2.172 &        7.25 &   0.000193  \\ 
1.10 &  1.0E-10  &    2.59E-05  &  -2.171 &        7.27 &   0.000188  \\ 
1.10 &  3.0E-10  &    2.41E-05  &  -2.168 &        7.32 &   0.000176  \\ 
1.10 &  1.0E-9  &    2.18E-05  &  -2.165 &        7.37 &   0.000161  \\ 
1.10 &  1.6E-9  &    2.07E-05  &  -2.164 &        7.39 &   0.000153  \\ 
1.10 &  3.0E-9  &    1.89E-05  &  -2.161 &        7.44 &   0.000141  \\ 
1.10 &  5.0E-9  &    1.74E-05  &  -2.159 &        7.47 &    0.00013  \\ 
1.10 &  1.0E-8  &    1.48E-05  &  -2.156 &        7.52 &   0.000111  \\ 
1.10 &  1.6E-8  &    1.29E-05  &  -2.152 &        7.59 &   9.77E-05  \\ 
1.10 &  3.0E-8  &    1.06E-05  &  -2.147 &        7.68 &   8.15E-05  \\ 
1.10 &  5.0E-8  &    8.94E-06  &  -2.142 &        7.77 &   6.95E-05  \\ 
1.10 &  1.0E-7  &    7.12E-06  &  -2.133 &        7.93 &   5.65E-05  \\ 
1.10 &  1.6E-7  &    6.14E-06  &  -2.125 &        8.08 &   4.96E-05  \\ 
1.20 &  1.0E-11  &    1.59E-05  &  -2.258 &        5.45 &   8.69E-05  \\ 
1.20 &  3.0E-11  &    1.41E-05  &  -2.257 &        5.47 &   7.68E-05  \\ 
1.20 &  5.0E-11  &    1.33E-05  &  -2.256 &        5.48 &    7.3E-05  \\ 
1.20 &  1.0E-10  &     1.3E-05  &  -2.255 &        5.49 &   7.13E-05  \\ 
1.20 &  3.0E-10  &     1.2E-05  &  -2.253 &        5.52 &   6.61E-05  \\ 
1.20 &  1.0E-9  &    1.09E-05  &  -2.250 &        5.55 &   6.04E-05  \\ 
1.20 &  1.6E-9  &    1.03E-05  &  -2.249 &        5.57 &   5.75E-05  \\ 
1.20 &  3.0E-9  &    9.44E-06  &  -2.247 &        5.59 &   5.28E-05  \\ 
1.20 &  5.0E-9  &    8.67E-06  &  -2.245 &        5.62 &   4.87E-05  \\ 
1.20 &  1.0E-8  &     7.4E-06  &  -2.242 &        5.66 &   4.19E-05  \\ 
1.20 &  1.6E-8  &    6.44E-06  &  -2.239 &         5.7 &   3.67E-05  \\ 
1.20 &  3.0E-8  &    5.31E-06  &  -2.235 &        5.75 &   3.05E-05  \\ 
1.20 &  5.0E-8  &    4.45E-06  &  -2.231 &         5.8 &   2.58E-05  \\ 
1.20 &  1.0E-7  &    3.52E-06  &  -2.224 &         5.9 &   2.08E-05  \\ 
1.20 &  1.6E-7  &       3E-06  &  -2.218 &        5.98 &   1.79E-05  \\ 
1.20 &  1.8E-7  &    2.88E-06  &  -2.216 &        6.01 &   1.73E-05  \\ 
1.25 &  1.0E-11  &    1.03E-05  &  -2.313 &        4.61 &   4.75E-05  \\ 
1.25 &  3.0E-11  &    8.98E-06  &  -2.312 &        4.62 &   4.15E-05  \\ 
1.25 &  5.0E-11  &    8.52E-06  &  -2.312 &        4.62 &   3.94E-05  \\ 
1.25 &  1.0E-10  &    8.29E-06  &  -2.311 &        4.63 &   3.84E-05  \\ 
1.25 &  3.0E-10  &    7.64E-06  &  -2.309 &        4.65 &   3.56E-05  \\ 
1.25 &  1.0E-9  &    6.93E-06  &  -2.306 &        4.69 &   3.25E-05  \\ 
1.25 &  1.6E-9  &    6.58E-06  &  -2.305 &         4.7 &   3.09E-05  \\ 
1.25 &  3.0E-9  &    6.02E-06  &  -2.303 &        4.72 &   2.84E-05  \\ 
1.25 &  5.0E-9  &    5.52E-06  &  -2.301 &        4.74 &   2.62E-05  \\ 
1.25 &  1.0E-8  &    4.71E-06  &  -2.298 &        4.77 &   2.25E-05  \\ 
1.25 &  1.6E-8  &    4.12E-06  &  -2.296 &         4.8 &   1.97E-05  \\ 
1.25 &  3.0E-8  &    3.39E-06  &  -2.292 &        4.84 &   1.64E-05  \\ 
1.25 &  5.0E-8  &    2.86E-06  &  -2.289 &        4.87 &   1.39E-05  \\ 
1.25 &  1.0E-7  &    2.24E-06  &  -2.282 &        4.95 &   1.11E-05  \\ 
1.25 &  1.6E-7  &    1.88E-06  &  -2.277 &        5.01 &   9.44E-06  \\ 
1.25 &  2.0E-7  &    1.74E-06  &  -2.274 &        5.05 &   8.79E-06  \\ 
1.30 &  1.0E-11  &    5.82E-06  &  -2.386 &        3.75 &   2.18E-05  \\ 
1.30 &  3.0E-11  &    5.06E-06  &  -2.385 &        3.76 &    1.9E-05  \\ 
1.30 &  5.0E-11  &    4.78E-06  &  -2.385 &        3.76 &    1.8E-05  \\ 
1.30 &  1.0E-10  &    4.63E-06  &  -2.384 &        3.77 &   1.74E-05  \\ 
1.30 &  3.0E-10  &    4.26E-06  &  -2.382 &        3.78 &   1.61E-05  \\ 
1.30 &  1.0E-9  &    3.88E-06  &  -2.380 &         3.8 &   1.47E-05  \\ 
1.30 &  1.5E-9  &    3.71E-06  &  -2.378 &        3.82 &   1.42E-05  \\ 
1.30 &  3.0E-9  &    3.37E-06  &  -2.376 &        3.84 &   1.29E-05  \\ 
1.30 &  1.0E-8  &    2.65E-06  &  -2.372 &        3.87 &   1.02E-05  \\ 
1.30 &  3.0E-8  &    1.92E-06  &  -2.366 &        3.93 &   7.52E-06  \\ 
1.30 &  1.0E-7  &    1.25E-06  &  -2.357 &        4.01 &   5.01E-06  \\ 
1.30 &  1.6E-7  &    1.06E-06  &  -2.352 &        4.05 &   4.29E-06  \\ 
1.30 &  2.0E-7  &    9.75E-07  &  -2.350 &        4.07 &   3.97E-06  \\ 
1.30 &  2.2E-7  &    9.43E-07  &  -2.348 &        4.09 &   3.86E-06  \\ 
1.31 &  1.0E-11  &    5.06E-06  &  -2.404 &        3.57 &   1.81E-05  \\ 
1.31 &  3.0E-11  &    4.38E-06  &  -2.403 &        3.58 &   1.57E-05  \\ 
1.31 &  5.0E-11  &    4.16E-06  &  -2.403 &        3.58 &   1.49E-05  \\ 
1.31 &  1.0E-10  &     3.9E-06  &  -2.402 &        3.59 &    1.4E-05  \\ 
1.31 &  3.0E-10  &    3.59E-06  &  -2.400 &         3.6 &   1.29E-05  \\ 
1.31 &  1.0E-9  &    3.25E-06  &  -2.398 &        3.62 &   1.18E-05  \\ 
1.31 &  3.0E-9  &    2.83E-06  &  -2.394 &        3.65 &   1.03E-05  \\ 
1.31 &  1.0E-8  &    2.24E-06  &  -2.390 &        3.69 &   8.27E-06  \\ 
1.31 &  3.0E-8  &    1.64E-06  &  -2.384 &        3.74 &   6.13E-06  \\ 
1.31 &  1.0E-7  &    1.09E-06  &  -2.375 &        3.82 &   4.14E-06  \\ 
1.31 &  2.0E-7  &    8.47E-07  &  -2.368 &        3.88 &   3.29E-06  \\ 
1.31 &  2.5E-7  &    7.81E-07  &  -2.365 &         3.9 &   3.05E-06  \\ 
1.31 &  2.6E-7  &    7.69E-07  &  -2.364 &        3.91 &   3.01E-06  \\ 
1.32 &  1.0E-11  &    4.35E-06  &  -2.424 &        3.38 &   1.47E-05  \\ 
1.32 &  3.0E-11  &    3.76E-06  &  -2.423 &        3.39 &   1.28E-05  \\ 
1.32 &  5.0E-11  &    3.57E-06  &  -2.422 &         3.4 &   1.21E-05  \\ 
1.32 &  1.0E-10  &    3.35E-06  &  -2.421 &        3.41 &   1.14E-05  \\ 
1.32 &  3.0E-10  &    3.08E-06  &  -2.420 &        3.41 &   1.05E-05  \\ 
1.32 &  1.0E-9  &    2.79E-06  &  -2.417 &        3.44 &    9.6E-06  \\ 
1.32 &  3.0E-9  &    2.43E-06  &  -2.414 &        3.46 &   8.41E-06  \\ 
1.32 &  1.0E-8  &    1.93E-06  &  -2.409 &         3.5 &   6.77E-06  \\ 
1.32 &  3.0E-8  &    1.42E-06  &  -2.404 &        3.54 &   5.01E-06  \\ 
1.32 &  1.0E-7  &    9.38E-07  &  -2.395 &        3.62 &   3.39E-06  \\ 
1.32 &  2.0E-7  &    7.31E-07  &  -2.388 &        3.67 &   2.69E-06  \\ 
1.32 &  2.5E-7  &    6.73E-07  &  -2.385 &         3.7 &   2.49E-06  \\ 
1.32 &  2.7E-7  &    6.54E-07  &  -2.384 &        3.71 &   2.42E-06  \\ 
1.33 &  1.0E-11  &    3.68E-06  &  -2.445 &         3.2 &   1.18E-05  \\ 
1.33 &  3.0E-11  &    3.18E-06  &  -2.444 &        3.21 &   1.02E-05  \\ 
1.33 &  5.0E-11  &    3.02E-06  &  -2.444 &        3.21 &   9.68E-06  \\ 
1.33 &  1.0E-10  &    2.83E-06  &  -2.443 &        3.21 &    9.1E-06  \\ 
1.33 &  3.0E-10  &    2.61E-06  &  -2.441 &        3.23 &   8.41E-06  \\ 
1.33 &  1.0E-9  &    2.37E-06  &  -2.439 &        3.24 &   7.68E-06  \\ 
1.33 &  3.0E-9  &    2.07E-06  &  -2.435 &        3.27 &   6.76E-06  \\ 
1.33 &  1.0E-8  &    1.64E-06  &  -2.431 &         3.3 &   5.43E-06  \\ 
1.33 &  3.0E-8  &    1.21E-06  &  -2.425 &        3.35 &   4.04E-06  \\ 
1.33 &  1.0E-7  &       8E-07  &  -2.417 &        3.41 &   2.73E-06  \\ 
1.33 &  2.0E-7  &    6.22E-07  &  -2.410 &        3.47 &   2.16E-06  \\ 
1.33 &  2.5E-7  &    5.73E-07  &  -2.407 &        3.49 &      2E-06  \\ 
1.33 &  2.7E-7  &    5.57E-07  &  -2.406 &         3.5 &   1.95E-06  \\ 
1.34 &  1.0E-11  &    3.05E-06  &  -2.469 &           3 &   9.18E-06  \\ 
1.34 &  3.0E-11  &    2.65E-06  &  -2.468 &        3.01 &   7.96E-06  \\ 
1.34 &  5.0E-11  &    2.51E-06  &  -2.467 &        3.02 &   7.57E-06  \\ 
1.34 &  1.0E-10  &    2.35E-06  &  -2.467 &        3.02 &    7.1E-06  \\ 
1.34 &  3.0E-10  &    2.17E-06  &  -2.465 &        3.03 &   6.57E-06  \\ 
1.34 &  1.0E-9  &    1.97E-06  &  -2.462 &        3.05 &   6.01E-06  \\ 
1.34 &  3.0E-9  &    1.72E-06  &  -2.459 &        3.07 &    5.3E-06  \\ 
1.34 &  1.0E-8  &    1.37E-06  &  -2.454 &        3.11 &   4.26E-06  \\ 
1.34 &  3.0E-8  &    1.01E-06  &  -2.449 &        3.15 &   3.18E-06  \\ 
1.34 &  1.0E-7  &     6.7E-07  &  -2.441 &         3.2 &   2.15E-06  \\ 
1.34 &  2.0E-7  &    5.18E-07  &  -2.434 &        3.26 &   1.69E-06  \\ 
1.34 &  2.5E-7  &    4.75E-07  &  -2.431 &        3.28 &   1.56E-06  \\ 
1.34 &  2.7E-7  &    4.61E-07  &  -2.430 &        3.29 &   1.52E-06  \\ 
1.34 &  3.0E-7  &    4.43E-07  &  -2.428 &         3.3 &   1.46E-06  \\ 
1.35 &  1.0E-11  &    2.48E-06  &  -2.496 &         2.8 &   6.94E-06  \\ 
1.35 &  3.0E-11  &    2.16E-06  &  -2.495 &        2.81 &   6.07E-06  \\ 
1.35 &  5.0E-11  &    2.06E-06  &  -2.494 &        2.82 &   5.79E-06  \\ 
1.35 &  1.0E-10  &    1.95E-06  &  -2.493 &        2.82 &   5.51E-06  \\ 
1.35 &  3.0E-10  &    1.81E-06  &  -2.492 &        2.83 &   5.13E-06  \\ 
1.35 &  1.0E-9  &    1.65E-06  &  -2.489 &        2.85 &   4.69E-06  \\ 
1.35 &  1.6E-9  &    1.56E-06  &  -2.488 &        2.85 &   4.46E-06  \\ 
1.35 &  3.0E-9  &    1.44E-06  &  -2.486 &        2.87 &   4.12E-06  \\ 
1.35 &  5.0E-9  &    1.32E-06  &  -2.484 &        2.88 &   3.81E-06  \\ 
1.35 &  1.0E-8  &    1.14E-06  &  -2.481 &         2.9 &    3.3E-06  \\ 
1.35 &  1.6E-8  &    1.01E-06  &  -2.478 &        2.92 &   2.95E-06  \\ 
1.35 &  3.0E-8  &    8.35E-07  &  -2.475 &        2.94 &   2.46E-06  \\ 
1.35 &  5.0E-8  &    7.05E-07  &  -2.472 &        2.96 &   2.09E-06  \\ 
1.35 &  1.0E-7  &    5.52E-07  &  -2.467 &           3 &   1.65E-06  \\ 
1.35 &  1.6E-7  &    4.64E-07  &  -2.463 &        3.02 &    1.4E-06  \\ 
1.35 &  2.0E-7  &    4.27E-07  &  -2.460 &        3.04 &    1.3E-06  \\ 
1.35 &  2.5E-7  &    3.94E-07  &  -2.458 &        3.06 &    1.2E-06  \\ 
1.36 &  1.0E-11  &    1.96E-06  &  -2.527 &        2.59 &   5.07E-06  \\ 
1.36 &  3.0E-11  &     1.7E-06  &  -2.526 &         2.6 &   4.41E-06  \\ 
1.36 &  5.0E-11  &    1.61E-06  &  -2.525 &         2.6 &    4.2E-06  \\ 
1.36 &  1.0E-10  &    1.51E-06  &  -2.524 &        2.61 &   3.94E-06  \\ 
1.36 &  3.0E-10  &     1.4E-06  &  -2.522 &        2.62 &   3.66E-06  \\ 
1.36 &  1.0E-9  &    1.27E-06  &  -2.520 &        2.63 &   3.35E-06  \\ 
1.36 &  3.0E-9  &    1.11E-06  &  -2.517 &        2.65 &   2.95E-06  \\ 
1.36 &  1.0E-8  &    8.91E-07  &  -2.511 &        2.69 &   2.39E-06  \\ 
1.36 &  3.0E-8  &    6.64E-07  &  -2.506 &        2.72 &    1.8E-06  \\ 
1.36 &  5.0E-8  &     5.6E-07  &  -2.502 &        2.74 &   1.54E-06  \\ 
1.36 &  1.0E-7  &    4.37E-07  &  -2.497 &        2.78 &   1.21E-06  \\ 
1.36 &  2.0E-7  &    3.38E-07  &  -2.491 &        2.81 &    9.5E-07  \\ 
1.36 &  2.5E-7  &     3.1E-07  &  -2.488 &        2.83 &   8.77E-07  \\ 
1.36 &  2.6E-7  &    3.05E-07  &  -2.488 &        2.83 &   8.64E-07  \\ 
1.37 &  1.0E-11  &    1.49E-06  &  -2.563 &        2.37 &   3.52E-06  \\ 
1.37 &  3.0E-11  &     1.3E-06  &  -2.562 &        2.37 &   3.08E-06  \\ 
1.37 &  5.0E-11  &    1.24E-06  &  -2.561 &        2.38 &   2.95E-06  \\ 
1.37 &  1.0E-10  &    1.19E-06  &  -2.560 &        2.38 &   2.83E-06  \\ 
1.37 &  3.0E-10  &    1.06E-06  &  -2.558 &        2.39 &   2.55E-06  \\ 
1.37 &  1.0E-9  &    9.68E-07  &  -2.556 &         2.4 &   2.33E-06  \\ 
1.37 &  3.0E-9  &    8.49E-07  &  -2.552 &        2.43 &   2.06E-06  \\ 
1.37 &  1.0E-8  &    6.84E-07  &  -2.547 &        2.46 &   1.68E-06  \\ 
1.37 &  3.0E-8  &    5.05E-07  &  -2.541 &        2.49 &   1.26E-06  \\ 
1.37 &  5.0E-8  &    4.27E-07  &  -2.538 &        2.51 &   1.07E-06  \\ 
1.37 &  1.0E-7  &    3.34E-07  &  -2.533 &        2.54 &   8.46E-07  \\ 
1.37 &  2.0E-7  &    2.57E-07  &  -2.527 &        2.57 &   6.59E-07  \\ 
1.37 &  2.5E-7  &    2.34E-07  &  -2.524 &        2.59 &   6.05E-07  \\ 
1.37 &  2.7E-7  &    2.27E-07  &  -2.523 &        2.59 &   5.88E-07  \\ 
1.37 &  3.0E-7  &    2.18E-07  &  -2.522 &         2.6 &   5.67E-07  \\ 
1.38 &  1.0E-11  &    1.06E-06  &  -2.607 &        2.12 &   2.25E-06  \\ 
1.38 &  3.0E-11  &    9.33E-07  &  -2.606 &        2.13 &   1.99E-06  \\ 
1.38 &  5.0E-11  &    8.99E-07  &  -2.605 &        2.13 &   1.92E-06  \\ 
1.38 &  1.0E-10  &     8.3E-07  &  -2.604 &        2.14 &   1.78E-06  \\ 
1.38 &  3.0E-10  &    7.72E-07  &  -2.602 &        2.15 &   1.66E-06  \\ 
1.38 &  1.0E-9  &    7.02E-07  &  -2.599 &        2.16 &   1.52E-06  \\ 
1.38 &  1.6E-9  &    6.68E-07  &  -2.598 &        2.17 &   1.45E-06  \\ 
1.38 &  3.0E-9  &    6.15E-07  &  -2.596 &        2.18 &   1.34E-06  \\ 
1.38 &  5.0E-9  &    5.69E-07  &  -2.593 &        2.19 &   1.25E-06  \\ 
1.38 &  1.0E-8  &    4.92E-07  &  -2.590 &        2.21 &   1.09E-06  \\ 
1.38 &  1.6E-8  &    4.38E-07  &  -2.587 &        2.22 &   9.74E-07  \\ 
1.38 &  3.0E-8  &    3.67E-07  &  -2.584 &        2.24 &   8.22E-07  \\ 
1.38 &  5.0E-8  &       3E-07  &  -2.581 &        2.25 &   6.77E-07  \\ 
1.38 &  1.0E-7  &    2.44E-07  &  -2.576 &        2.28 &   5.57E-07  \\ 
1.38 &  1.6E-7  &    2.05E-07  &  -2.572 &         2.3 &   4.73E-07  \\ 
1.38 &  2.0E-7  &     1.9E-07  &  -2.570 &        2.31 &   4.39E-07  \\ 
1.38 &  2.5E-7  &    1.75E-07  &  -2.567 &        2.33 &   4.07E-07  \\ 
1.38 &  2.7E-7  &     1.7E-07  &  -2.566 &        2.33 &   3.96E-07  
\enddata
%
\end{deluxetable}

\end{document}